\documentclass[pra,twocolumn,aps,showpacs,superscriptaddress]{revtex4}
\usepackage{epsfig}
\usepackage{bm}
\usepackage{amsmath}
\begin{document}

\author{E. Fratini}
\affiliation{The Abdus Salam International Centre for Theoretical Physics, 34151 Trieste, Italy}

\author{P. Pieri}
\affiliation{School of Science and Technology, Physics Division, University of Camerino and CNISM, I-62032 Camerino, Italy}
\affiliation{INFN, Sezione di Perugia, Perugia, Italy}

\title{Single-particle spectral functions in the normal phase of a strongly-attractive Bose-Fermi mixture}

\date{\today}

\begin{abstract}
We calculate the single-particle spectral functions and quasi-particle dispersions for a Bose-Fermi mixture when the boson-fermion attraction is sufficiently strong to suppress completely the condensation of bosons at zero temperature.  Within a T-matrix diagrammatic approach, we  
vary the boson-fermion attraction from the critical value where the boson condensate first disappears to the strongly attractive (molecular) regime and study the effect of both  mass- and density-imbalance on the spectral weights and dispersions.  An interesting spectrum of  particle-hole excitations mixing two different Fermi surfaces is found.  These unconventional excitations could be produced and explored experimentally with radio-frequency spectroscopy.
\end{abstract}

\pacs{03.75.Ss,03.75.Hh,32.30.Bv,74.20.-z}
\maketitle

\section{Introduction}\label{intro}

Recently, there has been an upsurge of interest on resonantly interacting Bose-Fermi mixtures, with several theoretical \cite{Pow05,Dic05,Sto05,Avd06,Bor08,Mar08,Wat08,Fra10,Yu11,Lud11,Song11,Fra12,Ber12}
and experimental \cite{Gun06,Osp06,Osp06b,Zir08,Ni08,Wu11,Wu12,Park12,Heo12} works being devoted to this subject. Calculations with different theoretical approaches  have shown that for sufficiently strong attraction, and a density of bosons smaller than the density of fermions, the boson condensation is completely suppressed even at zero temperature in favor of a phase with dominant molecular correlations \cite{Pow05,Dic05,Mar08,Fra10,Lud11,Ber12}. 
 
Such a zero-temperature normal phase is of interest to the present work. Some interesting features were found indeed in this phase in Ref.~\cite{Fra12}, like, for example, the presence of an empty in region in the bosonic momentum distribution at low momenta (see Fig.~7 of Ref.~\cite{Fra12}). 
This peculiar behavior, which occurs below a certain boson concentration,  results from the interplay between  molecular binding and the constraints in momentum space imposed by the Fermi statistics on the unpaired fermions and on the composite fermions (molecules). 

Aim of the present paper is then to characterize better such a normal phase with dominant molecular correlations, by a thorough study  of the spectral functions and associated quasi-particle dispersions and widths at zero temperature. 
We  believe indeed that in the near future the powerful technique of radio-frequency spectroscopy \cite{Reg03,Gup03} which lately has been applied extensively to probe strongly interacting ultracold Fermi gases, might be applied to investigate successfully also Bose-Fermi mixtures.
 This technique (especially when momentum-resolved)  gives access to the single-particle excitation spectrum, and has led to significant advances in the comprehension of fundamental phenomena in strongly interacting Fermi gases, such as pairing \cite{Chi04,Shi07,Schu08},  pseudogap \cite{Ste08,Gae10,Per11,Pie11}, and polaron physics \cite{Schi08,Schi09}. 

The analysis of the spectral functions of the present paper will be based on the analytic continuation to the real frequency axis of a T-matrix diagrammatic approach  formulated in the imaginary frequency axis, that we have used previously to characterize the phase diagram and thermodynamic quantities of a resonant Bose-Fermi mixture~\cite{Fra10,Fra12}. 
We will find in this way quite a rich spectrum of single-particle excitations.  In particular, we will see that the  removal/injection of a boson from/to the system  
produces  an interesting spectrum of  particle-hole excitations involving simultaneously two different Fermi surfaces (namely, the unpaired-fermion and composite-fermion Fermi surfaces). For both the fermionic and bosonic spectral functions we will also find, in specific region of momentum space, non-trivial single-particle excitations with an infinite lifetime.

The paper is organized as follows. In section \ref{formalism} we present the theoretical formalism and the main equations describing the physical quantities of interest in our paper.  The numerical results for the bosonic and fermionic spectral functions and derived quantities are reported in section \ref{results}.  Results will  be presented for different interaction strengths, density-imbalance, and mass ratios between the two components. Section \ref{conclusions} presents our conclusions, while the appendix contains some analytic calculations and expressions.

\section{Formalism}\label{formalism}
We consider a mixture of single-component fermions and bosons in the presence of a broad Fano-Feshbach resonance tuning the boson-fermion scattering length $a$. While a Fermi-Fermi $s$-wave scattering length is excluded by Pauli principle, some degree of repulsion between the bosons is known to be necessary for the mechanical stability of the system~\cite{Yu11,Ber12}.  We note however that the instability problems are much less important in the normal (molecular) phase of interest to the present paper~\cite{Yu11,Ber12}. In addition, the Bose-Bose repulsion will affect in a minor way the spectral properties of such a phase in comparison to the strong-attractive interaction which leads to the formation of molecules.  We thus do not include explicitly the boson-boson repulsion in our calculations.

We start  then from the following (grand-canonical) Hamiltonian:
\begin{eqnarray}
H&=&\sum_{s}\int\! d {\bf r} \psi^{\dagger}_s({\bf r})(-\frac{\nabla^2}{2 m_s}-\mu_s)
\psi_s({\bf r}) \nonumber\\
&+& v_0 \int\! d{\bf r} \psi^{\dagger}_{\rm B}({\bf r})\psi^{\dagger}_{\rm F}({\bf r})
\psi_{\rm F}({\bf r})\psi_{\rm B}({\bf r})
\label{hamiltonian},
\end{eqnarray}
where $s$=B,F indicates the boson and fermion atomic species, respectively  (we set $\hbar=k_{{\rm B}}=1$ throughout this paper). The operator $\psi^{\dagger}_s({\bf r})$ creates a particle of mass $m_s$ and chemical potential $\mu_s$ at spatial position ${\bf r}$. 

Like for two-component Fermi gases (see, e.g., \cite{Pie00}), the bare strength of the contact interaction $v_0$ is expressed in terms of the boson-fermion scattering length $a$ as:
\begin{equation}
\frac{1}{v_0}=\frac{m_r}{2\pi a}-\int \! \frac{d{\bf k}}{(2 \pi)^3} 
\frac{2 m_r}{{\bf k}^2}\;,
\end{equation}
where $m_r=m_{\rm B} m_{\rm F}/(m_{\rm B}+m_{\rm F})$ is the reduced mass of the boson-fermion system. 

A natural length scale for the many-body system is provided by the average interparticle distance $n^{-1/3}$ (where $n=n_{\rm B}+n_{\rm F}$ is the total particle-number density,  $n_{\rm B}$ and $n_{\rm F}$ being the individual boson and fermion particle-number density, respectively). We thus introduce a fictitious Fermi momentum of the system $k_{\rm F}\equiv (3 \pi^2 n)^{1/3}$ (as for an equivalent two-component Fermi gas with a density equal to the total density of the system), and use the dimensionless coupling parameter $g=(k_{\rm F} a)^{-1}$ to describe the strength of the interaction. 

In Refs.~\cite{Fra10,Fra12}, the thermodynamic properties of a Bose-Fermi mixture in the normal phase (i.e. above the condensation critical temperature) were studied within a T-matrix approximation for the (finite-temperature) self-energies.  The corresponding equations for the bosonic and fermionic self-energies $\Sigma_{\rm B}$ and $\Sigma_{\rm F}$ at finite temperature are:
\begin{eqnarray}\label{selfb}
\Sigma_{\rm B}({\bf k},\omega_{\nu})&=&-T\int\!\!\frac{d {\bf P}}{(2\pi)^{3}}\sum_{m}\Gamma({\bf P},\Omega_{m})\nonumber\\
&\times& G_{\rm F}^{0}({\bf P}-{\bf k},\Omega_m-\omega_{\nu})\\
\label{selff}
\Sigma_{\rm F}({\bf k},\omega_{n})&=&T\int\!\!\frac{d {\bf P}}{(2\pi)^{3}}\Gamma({\bf P},\Omega_{m})\nonumber\\
&\times&\sum_{m}G_{\rm B}^{0}({\bf P}-{\bf k},\Omega_m-\omega_{n})
\end{eqnarray}
where the many-body  T-matrix $\Gamma({\bf P},\Omega_{m})$ is given by
\begin{eqnarray}
&&\Gamma({\bf P}, \Omega_m)=- \left\{\frac{m_{r}}{2\pi a}+\int\!\!\frac{d{\bf p}}{(2\pi)^{3}}
\right.\nonumber\\
&&\times \left.\left[\frac{1-f(\xi^{\rm F}_{{\bf P}-{\bf p}})+b(\xi^{\rm B}_{{\bf p}})}{\xi^{\rm F}_{{\bf P}-{\bf p}}+\xi^{\rm B}_{\bf p}-i\Omega_{m}}
-\frac{2m_{r}}{{\bf p}^{2}} \right]\right\}^{-1}.
\label{gamma}
\end{eqnarray}
In the above expressions $\omega_{\nu}=2\pi\nu T$  and $\omega_{n}=(2n+1)\pi T$, $\Omega_m=(2m+1)\pi T$ 
are bosonic and fermionic Matsubara frequencies, respectively, ($\nu,n,m$ being integer numbers), while $f(x)$ and $b(x)$ are the Fermi and Bose distribution functions at temperature $T$, 
and $\xi^s_{\bf p}={\bf p}^2/2m_s-\mu_s$.

The self-energies (\ref{selfb}) and (\ref{selff}) determine the dressed Green's functions $G_{s}$ via the Dyson's equation $G_{s}^{-1}=G_{s}^{0 \; -1}-\Sigma_{s}$ (with the bare Green's functions given by $G_{\rm B,F}^0({\bf k}, \omega_{\nu,n})^{-1}=i\omega_{\nu,n}-\xi^{\rm B,F}_{\bf k}$). The dressed Green's functions $G_s$ allow to calculate the boson and fermion momentum distribution functions $n_{s}({\bf k})$ through the equations:
\begin{eqnarray}\label{nbq}
n_{\rm B}({\bf k})&=&- T \sum_{\nu}G_{\rm B}({\bf k},\omega_{\nu})\,e^{i\omega_{\nu} 0^+}\\
\label{nfk}
n_{\rm F}({\bf k})&=& T \sum_{n}G_{\rm F}({\bf k},\omega_{n})\,e^{i\omega_{n} 0^+}\,,
\end{eqnarray}
which in turn determine the boson and fermion number densities:
\begin{equation}\label{nbf}
n_{s}=\int\!\!\frac{d {\bf k}}{(2\pi)^{3}} n_{s}({\bf k}) .
\end{equation}
The inversion of the equations (\ref{nbf})  fully determines the boson and fermion chemical potentials at given densities and temperature, and therefore the thermodynamic properties of the Bose-Fermi mixture in the normal phase. In Refs.~\cite{Fra10} and \cite{Fra12} we presented the results for the chemical potentials, the phase diagram and the momentum distribution functions  obtained from the above equations (for  Bose-Fermi mixtures with equal and different masses, respectively).  It was shown in particular that the T-matrix  approximation recovers a reliable description of both weak- and strong-coupling limits of the Bose-Fermi attraction, and that it predicts the presence of a quantum phase transition between the condensed and the normal phase.  Above a certain critical coupling strength the previous set of equations for the normal phase remains then valid down to zero temperature.

\subsection{Analytic continuation to the real-frequency axis} 

 The calculation of the single-particle spectral functions was not considered in our previous works, which were formulated in the imaginary frequency axis.
 This calculation requires in fact the analytic continuation of the previous equations to the real frequency axis. The bosonic and  fermionic  spectral functions $A_{{\rm B}}$  and $A_{{\rm F}}$ are indeed defined in terms of the imaginary part of the retarded Green's functions $G_{{\rm B}}^{{\rm R}}$ and $G_{{\rm F}}^{{\rm R}}$, respectively:
\begin{equation}\label{spectral}
A_{s}({\bf k}, \omega)=-\frac{1}{\pi} \Im G_{s}^{{\rm R}}({\bf k}, \omega).
\end{equation}

In order to calculate the retarded Green's functions one starts from the spectral representation for the many-body T-matrix:
\begin{equation}\label{gammar}
\Gamma({\bf P}, \Omega_{m})=-\int_{-\infty}^{\infty}\frac{d{\omega'}}{\pi}\frac{\Im \Gamma^{{\rm R}}({\bf P},\omega')}{i\Omega_{m}-\omega'}
\end{equation}
where the retarded T-matrix  $\Gamma^{{\rm R}}$ is obtained from Eq.~(\ref{gamma}) with the replacement $i\Omega_{m}\rightarrow \omega'+i\eta$,
where $\eta$ is a positive infinitesimal quantity.\\

By inserting the spectral representation (\ref{gammar}) in the expressions (\ref{selfb}) and ({\ref{selff}) for the Matsubara self-energies, the sum over the internal Matsubara frequency $\Omega_m$ can be done analytically. The replacement $i\omega_{\nu,n}\to\omega+ i \eta$ then yields the retarded
 self-energies: 
\begin{eqnarray}\label{selfbr}
\Sigma_{{\rm B}}^{{\rm R}}(\bf{k},\omega)&=&\int\!\!\frac{d {\bf P}}{(2\pi)^{3}}\int_{-\infty}^{\infty}\!\frac{d{\omega'}}{\pi} \Im \Gamma^{{\rm R}}({\bf P},\omega')\nonumber\\
&\times&\frac{f(\xi^{{\rm F}}_{{\bf P}-{\bf k}})-f(\omega')}{\omega-\omega'+\xi^{{\rm F}}_{{\bf P}-{\bf k}}+i\eta}\\
\label{selffr}
\Sigma_{{\rm F}}^{{\rm R}}(\bf{k},\omega)&=&\int\!\!\frac{d {\bf P}}{(2\pi)^{3}}\int_{-\infty}^{\infty}\!\frac{d{\omega'}}{\pi} \Im \Gamma^{{\rm R}}({\bf P},\omega')\nonumber\\
&\times&\frac{b(\xi^{{\rm B}}_{{\bf P}-{\bf k}})+f(\omega')}{\omega-\omega'+\xi^{{\rm B}}_{{\bf P}-{\bf k}}+i\eta}.
\end{eqnarray}
The bosonic and the fermionic retarded Green's functions are then given by
\begin{equation}\label{spectralG}
G_{s}^{{\rm R}}({\bf k}, \omega)^{-1}=\omega-\xi^{s}_{\bf k}-\Sigma_{s}^{{\rm R}}({\bf k}, \omega)+ i \eta  .
\end{equation}
By considering the zero-temperature limit of Eqs.~(\ref{selfbr}) and (\ref{selffr}) one obtaines
\begin{eqnarray}\label{selfbrTz}
\Sigma_{{\rm B}}^{{\rm R}}(\bf{k},\omega)&=&\int\!\!\frac{d {\bf P}}{(2\pi)^{3}}\left[
\int_{-\infty}^{\infty}\!\frac{d{\omega'}}{\pi} \frac{\Theta(-\xi^{{\rm F}}_{{\bf P}-{\bf k}})\Im \Gamma^{{\rm R}}({\bf P},\omega')}{\omega-\omega'+\xi^{{\rm F}}_{{\bf P}-{\bf k}}+i\eta} \right.\nonumber\\
&&\left.-\int_{-\infty}^{0}\!\frac{d{\omega'}}{\pi}\frac{\Im \Gamma^{{\rm R}}({\bf P},\omega')}{\omega-\omega'+\xi^{{\rm F}}_{{\bf P}-{\bf k}}+i\eta}\right],\\
\label{selffrTz}
\Sigma_{{\rm F}}^{{\rm R}}(\bf{k},\omega)&=&\int\!\!\frac{d {\bf P}}{(2\pi)^{3}}\int_{-\infty}^{0}\!\frac{d{\omega'}}{\pi}\frac{\Im \Gamma^{{\rm R}}({\bf P},\omega')}{\omega-\omega'+\xi^{{\rm B}}_{{\bf P}-{\bf k}}+i\eta},
\end{eqnarray}
where it has been assumed $\mu_{\rm B} < 0$, as it always occurs in the normal phase.
The angular part of the integrals over the momentum $\bf P$ in the expressions (\ref{selfbrTz}) and (\ref{selffrTz}) can be calculated analytically. By taking then $\eta \rightarrow 0^{+}$, after some further manipulations, one is left with a two-dimensional integral to be calculated numerically (or just a one-dimensional integral for the contributions originating from the poles of
$\Gamma({\bf P},z)$ on the real axis). Details of these manipulations are reported in the appendix. 
The spectral functions are then determined by Eqs.~(\ref{spectral}) and (\ref{spectralG}). Note that the infinitesimal imaginary term 
$i\eta$ in the denominator of the retarded Green's function is relevant only when  $\Im\Sigma^{\rm R}_s({\bf k},\omega)=0$ and $\omega-\xi_{\bf k}^{s}-\Re \Sigma_{s}^{{\rm R}}({\bf k}, \omega)=0$, corresponding to fully undamped single-particle excitations,  yielding delta-like contributions to the spectral functions with weight  $\left \lvert 1-\frac{\partial}{\partial \omega} \Re \Sigma^{\rm R}_s(\bf k, \omega)\right \rvert^{-1}$.

Both boson and fermion spectral functions satisfy the sum rule 
\begin{equation}
\int_{-\infty}^{\infty}d{\omega} \, A_{{\rm B, F}}({\bf k}, \omega)=1 ;
\end{equation}
however, while $A_{{\rm F}}({\bf k}, \omega)$ is always positive,  
\begin{equation}
{\rm sgn}( \omega) A_{{\rm B}}({\bf k}, \omega)\ge 0 , 
\end{equation}
as it can be seen from the Lehmann representation~\cite{Fetter} of the two spectral functions which, for $T=0$, reads
\begin{equation}
A_s({\bf k}, \omega)=\begin{cases} \sum_n\vert\langle n\vert (c_{\bf k}^{s})^{\dagger}\vert\Psi_0\rangle\vert^2\delta(\omega-\epsilon_n^{N_{s}+1,N_{\bar{s} } })&  \omega >0\\ \sum_n\mp\vert\langle n\vert c_{\bf k}^{s}\vert\Psi_0\rangle\vert^2\delta(\omega+\epsilon_n^{N_{s}-1,N_{\bar{s} } })&  \omega < 0 ,
\label{lehmann}
\end{cases}
\end{equation}
where the minus (plus) sign in the above expression refers to the boson (fermion) case.  In Eq.~(\ref{lehmann}), the sum extends
over all the eigenstates $\vert n\rangle$ of $H$ which have a non-vanishing overlap with the state obtained by adding (subtracting) a particle of species $s$ with momentum ${\bf k}$ to (from) the 
ground-state $\vert\Psi_0\rangle$ of the system with  $N_s$ particles of the species $s$ and $N_{\bar s}$ of the other one. These states belong then to the Hilbert space
with $N_s \pm 1$ and  $N_{\bar s} $ particles of the two species,  and have excitation energies $\epsilon_n^{N_{s}\pm 1,N_{\bar{s} } }$  {\em relative} to the  ground states of the corresponding system with $N_s \pm 1$ and  $N_{\bar s} $ particles. 

Note that the negative sign of the boson spectral function for negative frequency guarantees the positivity of the boson momentum distribution function. One has indeed for 
the bosonic and fermionic momentum distribution functions $n_{{\rm B,F}}({\bf k})$:
\begin{eqnarray}\label{nbfk}
n_{{\rm B}}({\bf k})&=&\int_{-\infty}^{\infty}\!d \omega \, b(\omega) A_{{\rm B}}({\bf k}, \omega)\\
n_{{\rm F}}({\bf k})&=&\int_{-\infty}^{\infty}\!d \omega \, f(\omega) A_{{\rm F}}({\bf k}, \omega),
\end{eqnarray}
which at zero-temperature reduce to 
\begin{equation}\label{nbfktzero}
n_{{\rm B,F}}({\bf k})= \mp \int_{-\infty}^{0}\!d{\omega} \, A_{{\rm B,F}}(\bf k, \omega),
\end{equation}
where the minus (plus) sign refers to bosons (fermions).

\section{Numerical results}
\label{results}

We pass now to present the numerical results obtained on the basis of the formalism developed in the previous section. 
Preliminary to the calculation of the spectral functions is the determination of the boson and fermion chemical potentials for given values of the coupling strength $g$ and boson and fermion densities $n_{\rm B}$ and $n_{\rm F}$. At zero temperature this can be done by using either Eqs.~(\ref{nbfktzero})  or the zero-temperature limit of Eqs.~(\ref{nbq})-(\ref{nbf}), as we did in Ref.~\cite{Fra12}.  The two calculations are completely independent, because they make use of expressions obtained 
alternatively on the imaginary or real frequency axis. We have resorted to the formulation on the imaginary axis to determine the chemical potentials, while we have used the comparison for the momentum distribution functions calculated with the two different approaches as a stringent check of the numerical calculations and of the analytic continuation procedure.

\begin{figure}[t]
\epsfxsize=8cm
\epsfbox{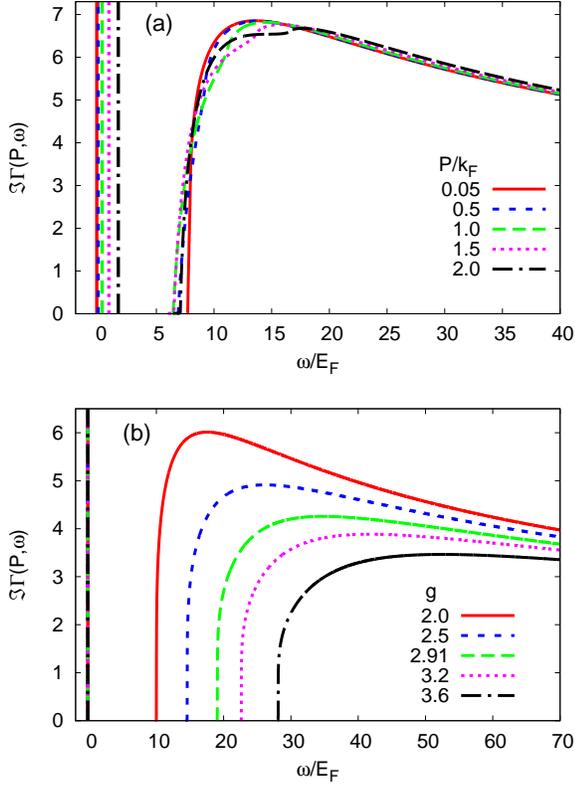}
\caption{(Color online) Composite-fermion spectral function $\Im \Gamma^{\rm R}({\bf P},\omega)$  [in units of $(2m_{\rm F} k_{\rm F})^{-1}$] vs.~frequency at zero temperature for a mixture with equal masses and density imbalance $(n_{{\rm F}}-n_{{\rm B}})/n=0.75$: (a) at $g_c=1.713$ for  different values of the momentum $P$ (with $\mu_{\rm B, F}(g_c)/E_{\rm F}=-6.421, 1.414)$;  (b) at $P=0$ for different couplings  (with corresponding values of the chemical potentials $\mu_{\rm B, F}/E_{\rm F}$=-8.615, 1.402; -13.2, 1.388;  -17.683, 1.377;  -21.259, 1.369; -26.736, 1.367, from top to bottom).}
\label{fig_gamma}
\end{figure}

\subsection{Retarded composite fermion propagator}
The many-body T-matrix $\Gamma$ recover the propagators of free molecules for sufficiently strong attraction (except for an overall normalization factor: $-2\pi/m_r^2 a$) \cite{Fra10}.  The imaginary part of the retarded T-matrix $\Gamma^{\rm R}({\bf P},\omega)$, which determines the retarded self-energies, is then essentially the spectral function of the composite-fermions (which become tightly bound molecules for strong attraction).
 This quantity is made 
of a delta-like
contribution, corresponding to the undamped motion of the composite-fermion,  plus a continuum part, which occurs above a threshold frequency $\omega_{\rm th}(P)$ such that the the decay channel 
(molecule $\to$ boson + fermion) is open. At finite temperature  $\omega_{\rm th}(P)={\bf P}^2/2 M - \mu_{\rm B} - \mu_{\rm F}$ 
(where $M=m_{\rm B}+m_{\rm F}$), as it can be readily derived from Eq.~(\ref{gamma}). 
The above expression for $\omega_{\rm th}(P)$ can be recast as $\omega_{\rm th}(P)=\xi^{\rm F}_{q}+\xi^{\rm B}_{P-q}$, with $q=m_{\rm F} P/M$, which emphasizes the final state after the decay.
[Here and in the following when the argument of the dispersion is a scalar, like  for $ \xi^{\rm B}_{P-q}$,  it should be interpreted in this way:  $\xi^{\rm B}_{P-q}=(P-q)^2/2 m_{\rm B} -\mu_{\rm B}$.] It is then clear that at zero temperature the complete Pauli blocking of fermions with wave vectors $q\le k_{\mu_{\rm F}}\equiv\sqrt{2 m_{\rm F}\mu_{\rm F}}$ due to the Fermi function appearing in  Eq.~(\ref{gamma}) (the Bose function being zero for negative $\mu_{\rm B}$) will 
shift the threshold frequency to
$\omega_{\rm th}(P)=\xi^{\rm F}_{k_{\mu_{\rm F}}}+\xi^{\rm B}_{P-k_{\mu_{\rm F}}}=\xi^{\rm B}_{P-k_{\mu_{\rm F}}}$  for $P < M  k_{\mu_{\rm F}}/m_{\rm F}$, while leaving unchanged  the previous expression of the threshold for $P \ge M k_{\mu_{\rm F}}/m_{\rm F}$.  

\begin{figure}[t]
\epsfxsize=8cm
\epsfbox{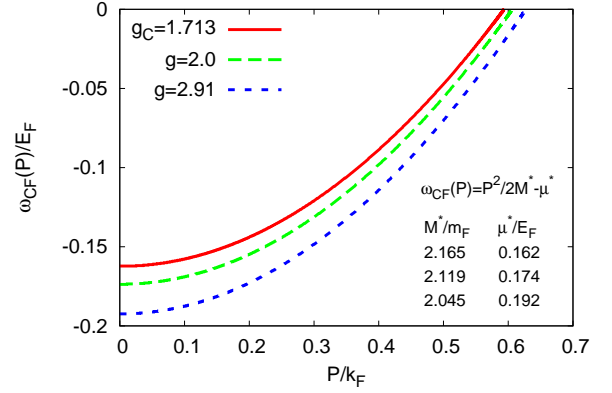}
\caption{(Color online) Dispersion  $\omega_{\rm CF}(P)$ of the pole of  $\Im \Gamma^{\rm R}({\bf P},\omega)$ at zero temperature for a mixture with equal masses and density imbalance $(n_{{\rm F}}-n_{{\rm B}})/n=0.75$
for three different coupling values ($g=g_c, g=2, g=2.9 $). }
\label{fig_gamma_pole}
\end{figure}

At zero temperature  the retarded T-matrix $\Gamma^{\rm R}({\bf P},\omega)$ can be expressed in a closed form  (reported in the appendix).
Here we show as an example in Fig.~\ref{fig_gamma}  the composite-fermion spectral function $\Im \Gamma^{\rm R}({\bf P},\omega)$ as a function of frequency for a mixture with equal masses and density imbalance $(n_{{\rm F}}-n_{{\rm B}})/n=0.75$ (at the critical coupling $g_c$ for different $P$,  and at $P=0$ for several couplings). 
One notices in Fig.~\ref{fig_gamma}(a) a rather weak dependence of the continuum part on $P$, while on Fig.~\ref{fig_gamma}(b) the main changes
in the continuum are accounted for by the coupling dependence of the chemical potential $\mu_{\rm B}$ and of the normalization factor $-2\pi/m_r^2 a$.

The position of the  delta-like contribution, which is signaled by the vertical lines in Fig.~\ref{fig_gamma},  is fitted extremely well by a quadratic dispersion  in the whole range of negative frequencies corresponding to the occupied states of the composite-fermions. As an example we report in Fig.~\ref{fig_gamma_pole} the dispersions $\omega_{\rm CF}(P)$ obtained numerically for three different coupling values, together with the associated fitting parameters
assuming a quadratic dispersion $\omega_{\rm CF}(P)=P^2/2M^*-\mu_{\rm CF}^*$ (we did not report the fitting curves because they coincide with the numerical data).
The dispersion $\omega_{\rm CF}(P)$ crosses zero at the wave-vector $k_{\rm CF}$ which defines the radius of the Fermi sphere of the composite fermions. Assuming the validity of the Luttinger's theorem for the composite-fermions, one would expect $k_{\rm CF}=(6\pi^2 n_{\rm B})^{1/3}$ when all bosons pair with fermions to form molecules. We have verified that the value $k_{\rm CF}=(6\pi^2 n_{\rm B})^{1/3}$ is reached only for sufficiently large $g$, while the value of $k_{\rm CF}$ at the critical coupling  corresponds, in the present case, to a density $n_{\rm CF}\equiv k_{\rm CF}^3/(6\pi^2)\simeq 0.85 n_{\rm B}$.  
Note further that at large values of $P$ the composite-fermion dispersion crosses over to a ``bare" dispersion with bare values $M^*=M$ and $\mu_{\rm CF}^*=\mu_{\rm B}+\mu_{\rm F} +\epsilon_0$ (a dispersion that holds for all values of $P$ in the strong-attraction limit).

Finally we have found that,  for all couplings of interest to the present paper, the weight $w(P)$ of the delta-like contribution of  $\frac{1}{\pi}\Im \Gamma^{\rm R}({\bf P},\omega)$ depends very weakly on $P$ and differs at most by a few percent from its strong-attraction limiting value $2\pi/m_r^2 a$.

\subsection{Spectral functions at the quantum critical point}
\begin{figure}[t]
\epsfxsize=8cm
\epsfbox{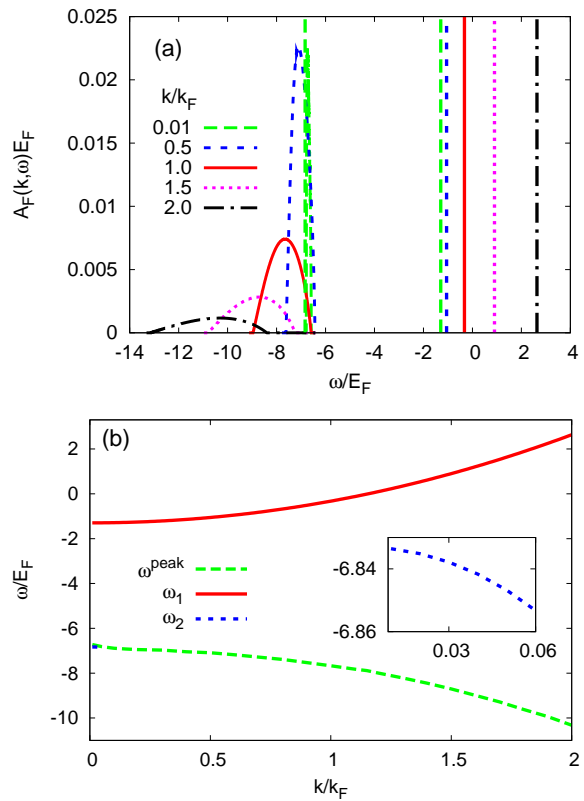}
\caption{(a) (Color online) Fermionic spectral function vs.~Êfrequency, for different values of the momentum $k$, for a mixture with equal masses and density imbalance $(n_{{\rm F}}-n_{{\rm B}})/n=0.75$ at the corresponding critical coupling $g_c=1.713$. (b) Dispersions $\omega_{1,2}(k)$  of the delta-like peaks and position of the broad peak 
$\omega_{\rm peak}$ of the fermionic spectral function for the same parameters as in  (a). The inset  is a detailed view of $ \omega_2(k)$.} 
\label{Afkom}
\end{figure}

We pass now to consider the single-particle spectral functions for  bosons and fermions, obtained by inserting the composite-fermion spectral function  
$\Im \Gamma^{\rm R}({\bf P},\omega)$ in Eqs.~(\ref{selffrTz})  and (\ref{selffrTz})  and then calculating the associated integrals over momenta and frequencies. As a first case we consider the same mixture of Fig.~\ref{fig_gamma}(a) with density imbalance $(n_{{\rm F}}-n_{{\rm B}})/n=0.75$ and equal masses. This is an example of a mixture for which the bosonic momentum distribution function was found~\cite{Fra12} to display the empty region at low momenta mentioned in the introduction.  
 Figures~\ref{Afkom} and \ref{Abkom}  presents the corresponding fermionic and bosonic spectral functions, respectively. 
Let us discuss first the fermionic spectral function $A_{\rm F}({\bf k},\omega)$, which is shown in Fig.~\ref{Afkom}(a)  for selected values of the momentum as a function of frequency. 
 One distinguishes clearly both delta-like peaks (represented by vertical lines) and broad peaks.  The broad peaks occur only at negative frequencies (recall that negative frequencies correspond to the removal of a particle) and are separated by a gap of the order of the molecular binding energy in vacuum  $\epsilon_0=1/(2 m_r a^2)$ from the delta-like peaks occurring at larger frequencies (in the present case $\epsilon_0=5.87 E_{\rm F}$). The broad peaks are associated then to excited states obtained by breaking 
 a molecule (by destruction of a fermion of momentum $k$). 
 The width of the broad peak increases with increasing momentum, similarly to what is observed with momentum-resolved radio-frequency spectroscopy in a strongly-interacting Fermi gas for the negative-frequency excitation branch~\cite{Per11}  (associated to pseudogap and pairing physics at low momenta and to the ``contact"  at large momenta~\cite{Schneider10}).
 The delta-like peaks occurring in Fig.~\ref{Afkom}(a) at frequencies $\omega\gtrsim -2 E_{\rm F}$,  absorb most of the spectral weight. Their dispersion $\omega_1(k)$ 
 crosses zero at the wave-vector $k_{\rm UF}$ where the momentum distribution function $n_{\rm F}(k)$ shows a jump. This momentum $k_{\rm UF}$ can be interpreted then as the radius of the Fermi sphere of the unpaired fermions. 
  We have also verified that the dispersion  $\omega_1(k)$  of Fig.~\ref{Afkom}(b)  is approximated very well  by a free dispersion $\frac{k^2}{2m_{{\rm F}}}-\mu_{{\rm UF}}$, where $\mu_{{\rm UF}}=k_{\rm UF}^2/(2 m_{\rm F})$.
 For negative frequencies (and thus $k < k_{\rm UF}$) this undamped single-particle excitation corresponds to the propagation of a hole in a filled Fermi sphere of unpaired fermions; for positive frequencies it corresponds to the propagation of an unpaired fermion added to the system with momentum $k > k_{\rm UF}$. Note finally that in a very small region of momenta close to zero ($k < 0.06 k_{\rm F}$ in Fig.~\ref{Afkom}(a)) a delta-like peak occurs also in the region of the spectrum associated with the breaking of a molecule. Its weight is small compared to the weight of the delta-like peak occurring at larger frequencies, but for $k\to 0$ it accounts essentially for all of the spectral weight associated with the ``molecular" part of the spectrum.   
 The dispersions of both delta-like peaks is reported in Fig.~{\ref{Afkom}(b). One can see that the secondary peak $\omega_2$ associated with the molecular branch merges rapidly into the broad peak, whose position  $\omega_{\rm peak}$ as a function of $k$ is also reported in  Fig.~{\ref{Afkom}(b). 
 The curve  $\omega_{\rm peak}$ is an inverted parabola, similarly to what occurs for the spectral function in the molecular limit of the BCS-BEC crossover~\cite{Ste08,Per11}. 
 
\begin{figure}[t]
\epsfxsize=8cm
\epsfbox{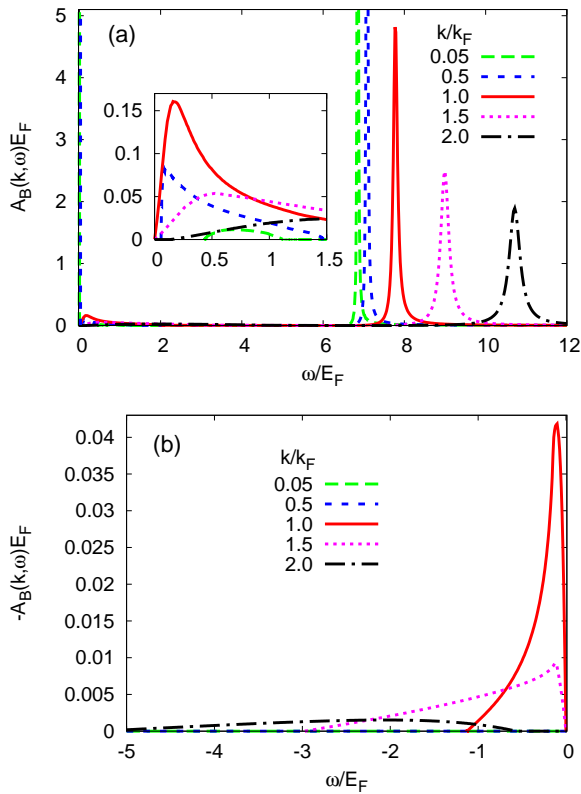}
\caption{(Color online) Bosonic spectral function  vs.~Êfrequency for different values of the momentum $k$, for a mixture with equal masses and density imbalance $(n_{{\rm F}}-n_{{\rm B}})/n=0.75$ at the corresponding critical coupling $g_c=1.713$. For clarity, the spectral function is presented separately for positive (a) and negative (b) frequencies.  The inset is a magnification of the spectral function at small positive frequencies (delta-like peaks are not shown in this case).}
\label{Abkom}
\end{figure}

The bosonic spectral function is reported in Fig.~\ref{Abkom} for the same parameters of Fig.~\ref{Afkom}. We observe  that in this case delta-like peaks are present only at positive frequencies, very close to $\omega=0$, and for momenta $k\lesssim 0.5 k_F$.  
For every value of the momentum two further features can be distinguished  in Fig.~\ref{Abkom}  (a) at positive frequencies: a broad feature at low frequencies and a narrow one at larger frequencies. They correspond to two different branches of excited states to which the ground state is coupled after adding a boson of momentum $k$. We interpret the lower branch as
a continuum of ``mixed" particle-hole excitations, namely,  the creation of a particle-like excitation outside the composite-fermion Fermi sphere plus the creation of a  hole inside the Fermi sphere of the unpaired fermions. Such excitations do not require to break a molecule and thus starts at zero energy. The second branch is instead a sharp quasi-particle excitation corresponding to the propagation of an unpaired boson. The added boson does not form a molecule and its dispersion is thus separated from zero frequency by a gap of the order of the binding energy.  It is the analogous in the present context of the so-called ``repulsive branch" in a strongly-attractive Fermi gas, that recently has been studied rather extensively in the context of itinerant ferromagnetism in  ultracold Fermi gases (see, e.g., \cite{Pil10,Pek11,Mas11}). We will see below that the dispersion of this quasi-particle peak follows essentially a free dispersion. 

At negative frequencies (panel (b)) one notices first of all the absence of any spectral weight at low momenta 
($k/k_{\rm F}=0.05$ and 0.5 in the figure). This implies a vanishing bosonic momentum distribution at low momenta, as it was mentioned in the introduction.
The radius of this empty region in the bosonic momentum distribution function is expected on physical grounds to be determined by the difference  $k_{\rm UF}- k_{\rm CF}$ between the Fermi momenta of the unpaired  and composite fermions \cite{Fra12}. We will see below that within our approach the radius of the empty region is actually given by $k_{\mu_{\rm F}}- k_{\rm CF}$, the difference between $k_{\mu_{\rm F}}$  and $k_{\rm UF}$ being however rather small.  For example, for the present case  $k_{\mu_{\rm F}}- k_{\rm CF}=0.595 k_{\rm F}$  to be compared with $k_{\rm UF}- k_{\rm CF}=0.56 k_{\rm F}$.

Past this region, a continuum of excitations appears. We interpret it again  as a continuum of ``mixed" particle-hole excitations, with the role of the two Fermi spheres exchanged with respect to the branch at positive frequencies. Indeed,  after the removal of a boson from a molecule,  the resulting state contains one less molecule and one more unpaired fermion. It couples then to a continuum of  ``mixed" particle-hole excitations of the system with 
$N_{\rm B}-1$ bosons, where now the hole is in the composite-fermion Fermi sphere, while the particle-like excitation is outside the unpaired-fermion Fermi sphere. 

\subsection{From the quantum critical point to the molecular limit}
In Fig.~\ref{Ag1713-291} we compare the spectral functions for the same mixture discussed above at the critical coupling  and in the strongly-attractive (molecular)  regime. For this comparison we use intensity plots, which are particularly useful to see the main features of the spectral functions.

One notices first of all that the fermionic spectral functions (left panels) have a simpler structure than the bosonic ones (right panels) and that, for both spectral functions, no qualitative changes occur when passing from the critical coupling (top) to the molecular limit (bottom).   
In particular, the main delta-like peak, associated with the unpaired fermions,  does not change appreciably with the coupling.  The main effects of interaction on the unpaired fermions are indeed due to the residual interaction with the composite-fermions, which scales to zero with $a$ in the molecular limit, and which is anyway small for the density-imbalance considered here, at which the boson density, and consequently the molecular density, are small. The molecular continuum in the fermionic spectral function, on the other hand,  is shifted down when the coupling increases.

In this respect, it can be seen by inspection of Eq.~(\ref{selffrTz}) for the fermionic self-energy at zero-temperature,
that the molecular continuum in the fermionic spectral function corresponds to the frequencies $\omega=\omega_{\rm CF}(P)-\xi^{\rm B}_{{\bf P}-{\bf k}}$, where ${\bf P}$ ranges within the Fermi sphere of the composite-fermions (i.e., $P \le  k_{\rm CF}$). It is easy to verify then, that $\omega$ is limited from below by $ \omega^{<}(k)=-\xi^{\rm B}_{k+k_{\rm CF}}$ and from above by
\begin{equation}
\omega^{>}(k)=
\begin{cases}
\frac{k^2}{2(M^*-m_{\rm B})}-\mu^*_{\rm CF}+\mu_{\rm B}& \phantom{aa}  \frac{k}{k_{\rm CF}}  < 1- \frac{m_{\rm B}}{M^*}\\
 -\xi^{\rm B}_{k-k_{\rm CF}} &   \phantom{aa}\frac{k}{k_{\rm CF}} \geq 1- \frac{m_{\rm B}}{M^*}\; .
\end{cases}
\end{equation}
One  can see easily from the above expressions of the threshold frequencies $\omega^{<}(k)$ and $\omega^{>}(k)$ that the width of the molecular continuum increases linearly with $k$ for $k > k_{\rm CF}(1-m_{\rm B}/M^*)$, where it
is given by $2 k k_{\rm CF} /m_{\rm B}$,  and that the top of the ``molecular band" is reached for $k=k_{\rm CF}$, where it equals $\mu_{\rm B}$ (explaining in this way  the down-shift of the molecular continuum  for increasing coupling, since $\mu_{\rm B}\simeq -\epsilon_0$ \cite{Fra10} becomes more and more negative ). At $k=0$ the width reduces to the quite small value
$\frac{k_{\rm CF}^2}{2 m_{\rm B}}(1-m_{\rm B}/M^*)$.
  
Finally, we notice that the secondary delta-like dispersion $\omega_2(k)$ disappears in the molecular limit (bottom-left panel of Fig.~\ref{Ag1713-291}).

\begin{figure}[t]
\epsfxsize=8.5cm
\epsfbox{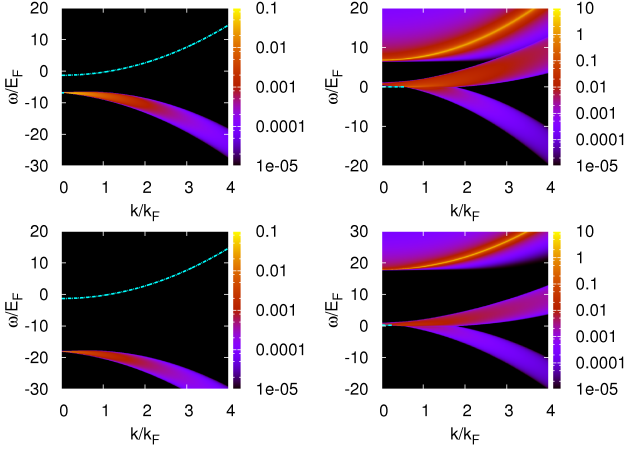}
\caption{(Color online) Intensity plots for the fermionic (left column) and bosonic (right column) spectral functions $A_{\rm F}({\bf k},\omega)$
and   $({\rm sgn} \omega) A_{\rm B}({\bf k},\omega)$ at zero temperature, for a mixture with density imbalance $(n_{{\rm F}}-n_{{\rm B}})/n=0.75$ and equal masses, at the critical coupling $g_c=1.713$ (top) and in the strong-attractive regime at $g=2.9 $ (bottom). Delta-like contributions are represented by  dashed-dotted lines. } 
\label{Ag1713-291}
\end{figure}

In the intensity plot for the bosonic spectral function,  one  sees at a glance the presence of  up to three different continuous frequency bands, depending  on 
the value of the wave-vector $k$.  One recognizes in the uppermost band the sharp quasi-particle dispersion, discussed above,  corresponding to the propagation of an unpaired boson added to the system. The sharp quasi-particle peak   can be fitted fairly well by a free dispersion $\omega=\xi_k^{\rm B} $.   
One can prove  also by starting from the expression  (\ref{selfbrTz}) for the bosonic self-energy, that the uppermost band, which originates from the continuum part of the composite-fermion spectral function $\Im \Gamma^{\rm R}({\bf P},\omega)$, is limited from below by $-\mu_{\rm B}$ for $k \le 2 k_{\mu_{\rm F}}$  and  by $\omega_{\rm th}(k-k_{\mu_{\rm F}})$ for $k > 2k_{\mu_{\rm F}}$ (before it merges with the intermediate band). These expressions show analytically that the uppermost band is separated by a gap  $\le -\mu_{\rm B}\simeq \epsilon_0$  from the ground state energy, as mentioned above.

The two remaining bands originate from the pole of the composite-fermion propagator, which contributes the following $\Im\Sigma_{\rm B}$ at zero temperature:

\begin{eqnarray}
\Im \Sigma_{\rm B}({\bf k},\omega)&=&\int\!\!\frac{d {\bf P}}{(2\pi)^{3}} w(P)\delta(\omega-\omega_{\rm CF}(P)+\xi^{{\rm F}}_{{\bf P}-{\bf k}})\nonumber\\
&\times&\left[\Theta(\omega)\Theta(-\xi^{{\rm F}}_{{\bf P}-{\bf k}})\Theta(\omega_{\rm CF}(P))\right. \nonumber\\
&&\left.-\Theta(-\omega)\Theta(\xi^{{\rm F}}_{{\bf P}-{\bf k}})\Theta(-\omega_{\rm CF}(P))\right]  ,
\label{enlight}
\end{eqnarray}
as it can be derived with some manipulations from Eq. ~(\ref{selfbrTz})  for the  delta-like contribution of $\Im \Gamma$. 

The above expression is particularly enlightening; one sees from it that the intermediate band in the intensity plots of the bosonic spectral function, which is located 
at positive frequencies, corresponds to a continuous of ``mixed" particle-hole processes, with a particle excitation at momentum ${\bf P}$ outside the composite-fermion Fermi sphere and a hole excitation at momentum  $ {\bf P}-{\bf k}$ inside the unpaired-fermion Fermi sphere. The lowermost band, which  occurs at negative frequencies, is again a ``mixed" particle-hole continuum where now the hole is at momentum ${\bf P}$ inside the composite-fermion Fermi sphere, while the particle excitation is at momentum  $ {\bf P}-{\bf k}$ outside the  Fermi sphere of the unpaired fermions. The two different kinds of mixed particle-hole excitations are excited, respectively, by adding/subtracting  a boson to/from the system. 
 
 Eq.~(\ref{enlight}) shows also  that at negative frequency, and when $k_{\rm CF} < k_{\mu_{\rm F}}$, $\Im \Sigma_{\rm B}=0$ for $k< k_{\mu_{\rm F}}-k_{\rm CF}$, since momentum conservation does not allow any  particle-hole process in this case.  The spectral-weight function  is thus identically zero at negative frequencies for these momenta,  thus leading via Eq.~(\ref{nbfktzero})  to a vanishing occupation of the corresponding momentum states. 
 This explains then the empty region in the bosonic momentum distribution function found in Ref.~\cite{Fra12}  and discussed in the introduction. Note that both $k_{\rm CF}$ and $k_{\mu_{\rm F}}$ depend essentially just on the density imbalance, the dependence on the coupling value being weak.  The corresponding absence of spectral weight at negative frequency and low momenta is clearly visible in both (right) panels. Note finally that we expect that a self-consistent calculation using  dressed Green's functions in the  self-energy would make   $k_{\rm UF}$  to appear in the place of $k_{\mu_{\rm F}}$ in the above expressions defining the empty region in the boson momentum distribution function. The difference between the two values is however in general rather small, in particular at large density imbalance where the fermionic self-energy is small.
  
From Eq.~(\ref{enlight}) one can also derive the threshold frequencies $\omega_{+}^{>}(k)$ and  $\omega_{+}^{<}(k)$ delimiting the intermediate band from above and below, respectively, as well as the corresponding frequencies  $\omega_{-}^{>}(k)$ and  $\omega_{-}^{<}(k)$ for the lowermost band.  We report their expression in the appendix. From those expressions one can see that at positive frequencies and for   $k_{\rm CF} < k_{\mu_{\rm F}}$, there is a small pocket close to zero frequency (specifically, for $0<\omega< \omega_{\rm CF}(k-k_{\mu_{\rm F}})$) where $\Im\Sigma_{{\rm B}}$ vanishes. Within this region one can observe in Fig.~\ref{Ag1713-291} the appearance of a delta-like peak, with a rather flat dispersion (as it was already noticed in Fig.~\ref{Abkom}).

\subsection{Changing the density imbalance and mass ratio}
\begin{figure}[t]
\epsfxsize=8.5cm
\epsfbox{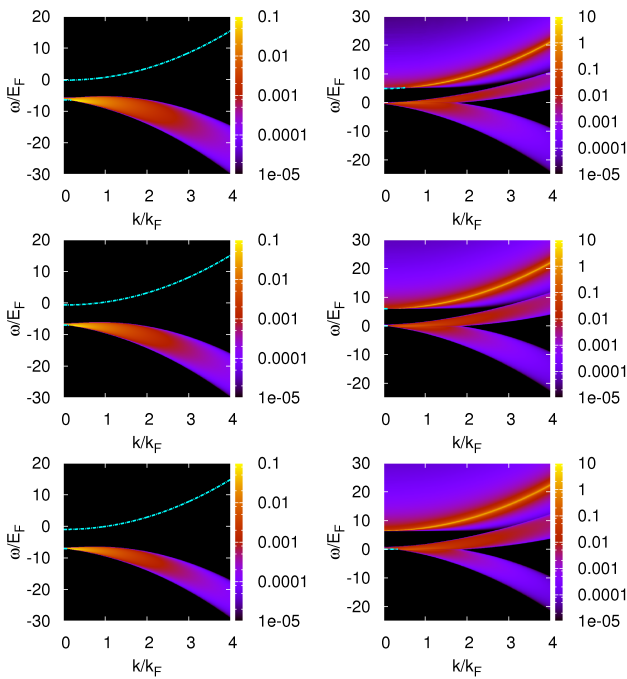}
\caption{(Color online) Intensity plots for the fermionic (left column) and bosonic (right column) spectral functions  $A_{\rm F}({\bf k},\omega)$
and   $({\rm sgn} \omega) A_{\rm B}({\bf k},\omega)$ at zero temperature, for mixtures with equal masses $m_{{\rm B}}/m_{{\rm F}}=1$ and different values of the density imbalance at the critical coupling $g_c$.  From  top to bottom: $(n_{{\rm F}}-n_{{\rm B}})/n=0$, $0.25$, $0.5$, with corresponding values of  $g_c=1.62$, $1.705$, $1.728$ (and chemical potentials $\mu_{\rm B, F}/E_{\rm F}=-5.186, 0.621; -6.010, 0.972; -6.357, 1.213$, respectively). Delta-like contributions are represented by  dashed-dotted lines.}
\label{Abf_imb}
\end{figure}
We finally analyze the dependence of the spectral functions at the critical coupling on the density imbalance and on the mass ratio $m_{\rm B}/m_{\rm F}$.  In Fig.~\ref{Abf_imb} we present then 
the intensity plots for three different density imbalances for $m_{{\rm B}}/m_{{\rm F}}=1$, while in Fig.~\ref{Abf_mass} we consider two different mass ratios 
at a fixed value (0.75) of the density imbalance.

We see from  Fig.~\ref{Abf_imb} that varying the density imbalance produces relatively minor effects on both spectral functions. In particular, in the fermionic spectral function the main visible effects are the shrinking of the width of the molecular band and the downward shift of the delta-like dispersion when the density imbalance is increased.  These effects are explained by the decrease of the composite-fermion Fermi  momentum $k_{\rm CF}$ (which controls the width of the molecular band) and the corresponding increase of the unpaired fermion Fermi momentum $k_{\rm UF}$ (which controls the delta-like dispersion $\omega_1(k)$) when the number of bosons decreases.  One has specifically $k_{\rm CF}/k_{\rm F}=0.97,0.87,0.75,0.59$ and $k_{\rm UF}/k_{\rm F}=0.46,0.84,1.02,1.15$ for density imbalance equal 0, 0.25, 0.5, and 0.75 (shown previously), respectively.

Note that the dispersion $\omega_1(k)$ of the unpaired fermions, which at large density imbalance is  essentially free, is somewhat renormalized by interaction for small density imbalance. As an example, for $n_{\rm F}=n_{\rm B}$ and $k\lesssim k_{\rm F}$ one can fit  quite accurately $\omega_1(k)$ with the expression $(k^2 -k_{\rm UF}^2)/2 m_{\rm F}^*$ with $m_{\rm F}^*=1.1 m_{\rm F}$, while at large wave-vectors, $\omega_1(k)$ crosses over to a free dispersion, since interaction effects are suppressed when a particle is added with large kinetic energy~\cite{Nis12}. 

Note further that, having defined $n_{\rm UF}\equiv k_{\rm UF}^3/6\pi^2$ and $n_{\rm CF}\equiv k_{\rm CF}^3/6\pi^2$, the equation  $n_{\rm UF}+n_{\rm CF}=n_{\rm F}$ is verified with great accuracy (i.e. within less than  0.5\%), indicating the validity of the Luttinger's theorem  for the total number of fermions 
(cf.~the discussion in  \cite{Pow05} on this point, albeit within a three-component model for a resonant Bose-Fermi mixture).
The corresponding values of $n_{\rm UF}$ and $n_{\rm CF}$, on the other hand, would suggest that  at the critical coupling a fraction of fermions and bosons do not contribute to the composite-fermion density, with  $n_{\rm CF}/ n_{\rm B} =$ 0.85 - 0.95, as it was noted already above for  the case $(n_{{\rm F}}-n_{{\rm B}})/n=0.75$.
It is not clear, however, to what extent this is jus a matter of definition of a quantity, such as the composite-fermion density, which becomes unambiguously defined only in the extreme molecular limit  $g\gg 1$.

\begin{center}
\begin{figure}[t]
\epsfxsize=8.5cm
\epsfbox{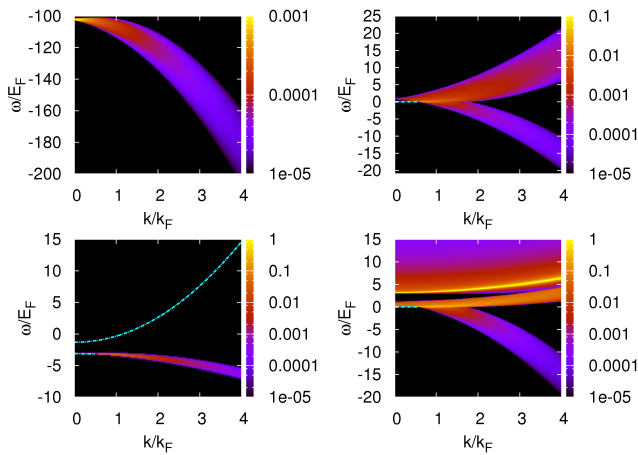}
\caption{(Color online) Intensity plots for the fermionic (left column) and bosonic (right column) spectral functions  $A_{\rm F}({\bf k},\omega)$
and  $({\rm sgn}  \omega) A_{\rm B}({\bf k},\omega)$ at zero temperature at the critical coupling $g_c$,  for mixtures with fixed density imbalance $(n_{{\rm F}}-n_{{\rm B}})/n=0.75$ and two different mass ratios: $m_{{\rm B}}/m_{{\rm F}}=0.2$ (top), $m_{{\rm B}}/m_{{\rm F}}=5.0$ (bottom), with corresponding values of  $g_c=4.1$, $1.325$
(and chemical potentials $\mu_{\rm B, F}/E_{\rm F}=-100.971, 1.455; -2.955, 1.395$, respectively). Delta-like contributions are represented by  dashed-dotted lines. For the case $m_{{\rm B}}/m_{{\rm F}}=0.2$ we focus on the molecular band of the fermionic spectrum (top left panel) and on the intermediate and lower band of the bosonic spectrum (top right panel).}
\label{Abf_mass}
\end{figure}
\end{center}

In the bosonic intensity plots (right column  of Fig.~\ref{Abf_imb}), one notices for the density-balanced case  the absence of the region without spectral weight at negative frequencies. This is consistent with our previous discussion of the depleted region, since in this case one has  $k_{\rm CF} > k_{\mu_{\rm F}}$ (and $k_{\rm CF}> k_{\rm UF}$ as well). One sees also that when the density-imbalance increases, the widths of the intermediate and lower band increase and decrease, respectively. Indeed, by examining the expressions for the threshold frequencies reported in the appendix, it can be easily seen that the two widths (for $k \ge k_{\mu_{\rm F}}+k_{\rm CF}$) are given by $2 k k_{\mu_{\rm F}}/M^*$ and $2 k k_{\rm CF}/m_{\rm F}$, respectively. The dependence of $ k_{\mu_{\rm F}}$ and $k_{\rm CF}$ on the density imbalance control then the two widths. Another interesting feature is  the behavior of the delta-like poles. We see that for a density balanced mixture, a delta-like dispersion
appears at low momenta just below the threshold of the upper band. This pole then enters into the upper band for $k\simeq 0.5 \, k_F$, where it becomes a narrow quasi-particle peak. We thus see that for a density-balanced mixture an unpaired boson with low momentum is fully protected from  decay.
When the density-imbalance increases, the delta-like dispersion is instead progressively transferred from the upper band to the intermediate band at small positive frequencies. 
 
 Figure~\ref{Abf_mass} shows that changing the mass ratio  $m_{{\rm B}}/m_{{\rm F}}$ produces quite a large change in both fermionic and bosonic spectral functions.
In particular,  a comparison with the case with equal masses reported in Fig.~\ref{Ag1713-291} shows that 
the ``molecular" branch of the fermionic spectrum is pulled down considerably when the boson mass is reduced  (top left panel), while it is slightly pushed up when the boson mass in increased to $m_{\rm B}= 5 m_{\rm F}$ (bottom left panel). Recalling that the energy scale for this branch is set by the binding energy $\epsilon_0$, this behavior is then explained  by the corresponding behavior of the critical  coupling $g_c$ (and thus  of the binding energy 
$\epsilon_0$) as a function of the mass ratio. It was shown indeed in Ref.~\cite{Fra12} that $g_c$ increases rapidly when $m_{\rm B}$ gets smaller than $m_{\rm F}$, while it first decreases and then slowly increases when $m_{\rm B}$ gets larger and larger compared to  $m_{\rm F}$ (with a minimum value of $g_c$  occurring for $m_{{\rm B}}/m_{{\rm F}}\simeq 5$). 

The width of the molecular band is also strongly affected by the mass ratio.  Indeed,  we have seen previously that this width is given by $2 k k_{\rm CF} /m_{\rm B}$, and thus when expressed in units of $E_{\rm F}$ is inversely proportional to the mass ratio  $m_{{\rm B}}/m_{{\rm F}}$.
Regarding the delta-like dispersions, the main one, corresponding to the propagation of an unpaired fermion,  is essentially unchanged when the boson mass is varied, since this dispersion  depends  on the unpaired-fermion effective mass and Fermi wave-vector, which depend essentially just on the density imbalance. The secondary delta-like peak,  which for equal masses  was confined to very small  momenta, extends its presence to larger momenta for $m_{{\rm B}}= 5 m_{{\rm F}}$, while it  is completely absent for  $m_{{\rm B}}= 0.2 m_{{\rm F}}$.\\

In the bosonic intensity plots (right column  of Fig.~\ref{Abf_mass}) one notices first all that the intermediate band  flattens and reduces its width when the boson mass increases. This is because this band of mixed particle-hole excitations is controlled essentially by the dispersion $\omega_{\rm CF}(P)$ of the composite-fermions, and  in particular by the effective mass $M^*\simeq m_{\rm B}+ m_{\rm F}$, which increases with $m_{\rm B}$, thus making the dispersion flatter and the width $2 k k_{\rm \mu_{\rm F}}/M^*$ smaller. The lower band, with holes in the composite-fermion Fermi sphere and particles outside the unpaired-fermion Fermi sphere is instead controlled essentially by the fermion dispersion $\xi^{\rm F}_k$ and thus does not change with $m_{\rm B}$. Finally, the uppermost band, associated with the propagation of an unpaired boson, is pushed to very large energies for $m_{\rm B}=0.2 m_{\rm F}$ (outside the range reported in Fig.~\ref{Abf_mass}), and  pulled down for $m_{\rm B}=5 m_{\rm F}$, following the behavior of the binding energy at $g_c$  vs.~$m_{\rm B}/m_{\rm F}$ discussed above. 
No significant effects are instead observed in the delta-like dispersion at small momenta.

\section{Concluding Remarks}\label{conclusions}

In this paper, we have presented a thorough study of the boson and fermion single-particle spectral functions in the zero-temperature normal phase of  a strongly-attractive Bose-Fermi mixture. This phase occurs in a mixture with $n_{\rm B}\le n_{\rm F}$ when the  pairing correlations induced by the boson-fermion attraction are sufficiently strong to suppress completely the boson condensate even at zero temperature. 
 We have found that the spectrum of single-particle excitations that can be extracted from the two spectral functions $A_{\rm B,F}({\bf k},\omega)$ is quite rich. The fermionic spectral function contains two main branches of excitations: (i) the propagation without damping of an unpaired fermion (or, at negative frequencies, of a hole in the unpaired fermion Fermi sphere); (ii) the ``molecular" dissociation spectrum. Interestingly, such a  dissociation occurs also when the fermion is removed from the unpaired-fermion Fermi sphere (i.e., for $k<k_{\rm UF}$), provided that an excitation energy of the order of the binding energy is furnished to the system.   This is because if a hole is produced in the unpaired-fermion Fermi sphere, the corresponding momentum state becomes available for pairing.  There are then exact eigenstates of the system where the hole is filled partially with fermions paired with bosons. The destruction of a fermion at momentum $k< k_{\rm UF}$ couples then also to the dissociation spectrum of  these excited states, where the momentum state $k$ is partially occupied by a fermion participating to pairing.   

The bosonic spectral function displays an even richer structure. Probably the most interesting feature that we have found in this case, is the presence of a spectrum of    ``mixed"  particle-hole excitations, namely, particle-hole excitations involving simultaneously two different Fermi surfaces.
These ``mixed'' particle-hole excitations can be produced either by removing a boson or by adding it to the system.  In the first case, the excitation is of the 
particle type for the unpaired fermions and of the hole type for the composite fermions, while in the second case the reverse occurs.
These unconventional excitations could be explored experimentally by momentum-resolved radio-frequency spectroscopy.
 Specifically, the branch associated with the removal of  a boson could be explored by applying a radio-frequency signal tuned about the transition frequency between two different hyperfine levels of the bosonic atoms. The exploration of the branch associated to the creation of a boson should instead rely on inverse radio-frequency spectroscopy. This would require to have in the trap also  a small population of  bosons in  a different hyperfine level with respect  to the main hyperfine level 
 involved in the Fano-Feshbach resonance tuning the boson-fermion interaction. The application of a radio-frequency signal would thus induce transitions from the  weakly populated level to the main one. Within linear response, the ensuing spectra (when momentum-resolved) would then be proportional to the part of the spectral function  associated with the creation of a particle (namely, the positive-frequency part of the spectrum).
 
 The spectrum at positive frequency contains finally two further features: a sharp quasi-particle excitation associated with the propagation of an unpaired boson added to the system, and an undamped  (delta-like) excitation which occurs only at small momenta and, depending on the density imbalance,  can appear just below the threshold of the mixed particle-hole  or  of the unpaired-boson branches of the spectrum. We do not have a clear interpretation for such an undamped excitation, which is however a robust feature within our theoretical approach, to the extent that  it was found for all the different  cases considered through our paper. 
 
 It is natural, finally,  to wonder about the validity of our results beyond the specific choice for the boson and fermion self-energies that we have made in our work.  
 We know already that the T-matrix choice for the self-energies becomes asymptotically exact in the extreme molecular limit  ($g \gg 1$), where it describes the correct limiting situation of a mixture of noninteracting molecules and unpaired fermions~\cite{Fra10,Fra12}. In the less extreme limit, the T-matrix self-energies tend to overestimate the effective repulsion between composite-fermions and unpaired fermions (as evidenced by the larger value of the molecule-fermion scattering length when compared with the exact value obtained from three-body calculations~\cite{sko57,Isk07}).
  This will have an impact on the numerical values of parameters like the effective masses of the composite and unpaired fermions, or the value of the critical coupling $g_c$. It will however leave essentially unchanged all the remaining main features, which depend just on the binding energy, degree of density imbalance and mass ratios.  
\appendix
\section{}
\subsection{Retarded self-energies}
We report here the analytic calculations leading to the final expressions for the bosonic and fermionic retarded self-energies, that we used for the numerical calculation of the spectral weight functions. 
Starting from Eq. ~(\ref{selffrTz}) and by doing an integration over the angles between ${\bf P}$ and ${\bf k}$  one obtains for the fermionic self-energy 

\begin{eqnarray}\label{selfflog}
\Sigma_{{\rm F}}^{{\rm R}}(\bf{k},\omega)&=&-\frac{m_{\rm B}}{k}\!\!\int_{0}^{+\infty}\! \frac{d P P}{4\pi^{2}}\int_{-\infty}^{0}\!d{\omega'}\frac{1}{\pi}\Im \Gamma^{{\rm R}}({\bf P},\omega')
\nonumber\\&\times&
\ln{\left[ \frac{\omega-\omega'+\xi^{\rm B}_{P-k}+i\eta}{\omega-\omega'+\xi^{\rm B}_{P+k}+i\eta}\right] }.
\end{eqnarray}
From this expression we see that only the delta-like contribution of  $\frac{1}{\pi}\Im \Gamma^{{\rm R}}$ contributes to the fermionic self-energy since the threshold frequency 
$\omega_{\rm th}(P)$ of the continuous part is always positive.  We recall in fact that
\begin{equation}
\omega_{\rm th}(P)=\begin{cases}\xi^{\rm B}_{P-k_{\mu_{\rm F}}}& P \le M  k_{\mu_{\rm F}}/m_{\rm F}\\
{\bf P}^2/2 M - \mu_{\rm B}-\mu_{\rm F}& P > M  k_{\mu_{\rm F}}/m_{\rm F},
\end{cases}
\end{equation}
and in the normal phase at $T=0$, $\mu_{\rm B}+\mu_{\rm F}\simeq -\epsilon_0$.

We have then 
\begin{equation}\label{selfffin}
\Sigma_{{\rm F}}^{{\rm R}}({\bf k},\omega)=-\frac{m_{\rm B}}{k}\!\!\!\int_{0}^{k_{\rm CF}}\!\!\frac{d P P}{4\pi^2} \, w(P) \left[ \ln{\vert \rho_{{\rm F}} \vert}+i\pi \Theta(-\rho_F)\right] 
\end{equation}
where $\rho_{{\rm F}}=(\omega-\omega_{\rm CF}(P)+\xi^{\rm B}_{P-k})/(\omega-\omega_{\rm CF}(P)+\xi^{\rm B}_{P+k})$,
and we have taken the limit $\eta\to 0^+$.
The bosonic self-energy is more complicated. We treat separately the two contributions appearing in Eq.~(\ref{selfbrTz}) and define $\Sigma_{{\rm B}}^{{\rm R}}({\bf k},\omega)=\Sigma_{{\rm B}}^{{\rm R}}({\bf k },\omega)^{\rm I}+\Sigma_{{\rm B}}^{{\rm R}}({\bf k },\omega)^{\rm II}$ with
\begin{eqnarray}\label{selfbrTz1}
\Sigma_{{\rm B}}^{{\rm R}}({\bf k},\omega)^{\rm I}=\int\!\!\frac{d {\bf P}}{(2\pi)^{3}}\!
\int_{-\infty}^{\infty}\!\!\!\frac{d{\omega'}}{\pi} \frac{\Theta(-\xi^{{\rm F}}_{{\bf P}-{\bf k}})\Im \Gamma^{{\rm R}}({\bf P},\omega')}{\omega-\omega'+\xi^{{\rm F}}_{{\bf P}-{\bf k}}+i\eta}\phantom{aa}&&\\
\label{selfbrTz2}
\Sigma_{{\rm B}}^{{\rm R}}({\bf k },\omega)^{\rm II}=-\int\!\!\frac{d {\bf P}}{(2\pi)^{3}}\!\int_{-\infty}^{0}\!\!\!\frac{d{\omega'}}{\pi}\frac{\Im \Gamma^{{\rm R}}({\bf P},\omega')}{\omega-\omega'+\xi^{{\rm F}}_{{\bf P}-{\bf k}}+i\eta}.\phantom{aa}&&
\end{eqnarray}
The term $\Sigma_{{\rm B}}^{{\rm R}}({\bf k },\omega)^{\rm II}$ is completely analogous to the fermionic self-energy and leads then to the expression
\begin{equation}
\Sigma_{{\rm B}}^{{\rm R}}({\bf k},\omega)^{\rm II}=\frac{m_{{\rm F}}}{k}\int_{0}^{k_{\rm CF}}\!\! \frac{d P P}{4\pi^2} \, w(P) \left[ \ln{\vert \rho_{{\rm B}} \vert}+i\pi \Theta(-\rho_B)\right] 
\end{equation}
where $\rho_{{\rm B}}=(\omega-\omega_{\rm CF}(P)+\xi^{\rm F}_{P-k})/(\omega-\omega_{\rm CF}(P)+\xi^{\rm F}_{P+k})$.
For $k_{\mu_{\rm F}}<k$ we have:
\begin{widetext}
\begin{equation}
\Sigma_{\rm B}^{\rm R}({\bf k},\omega)^{\rm I}=-\frac{m_{\rm F}}{k}  \!\!\!
\int_{k- k_{\mu_{\rm F}}}^{k+k_{\mu_{\rm F}}}\!\!\frac{dP P}{4\pi^{2}}\!\left\{\left[ \ln\vert \rho_{\rm B} \vert+i\pi \Theta(-\rho_{\rm B})\right] w(P)
+\int_{\omega_{\rm th}(P)}^{+\infty}\!\!\frac{d\omega'}{\pi}\Im \Gamma^{\rm R}({\bf P},\omega')\left[\ln \vert \rho_{\rm B}^{\rm I}\vert+i\pi \Theta(-\rho_{\rm B}^{\rm I})\right]\right\},
\end{equation}
where $\rho_{{\rm B}}^{\rm I}=(\omega-\omega'+\xi^{\rm F}_{P-k})/(\omega-\omega')$, while for $k_{\mu_{{\rm F}}}>k$ we have: 
\begin{eqnarray}\label{selfb14fin}
&&\Sigma_{{\rm B}}^{{\rm R}}({\bf k},\omega)^{\rm I}=\frac{m_{\rm F}}{k}\!\!\!\int_{0}^{k_{\mu_{{\rm F}}}-k}\!\frac{d P P}{4\pi^{2}} \left\{w(P)\left[ \ln{\vert \rho_{{\rm B}}\vert}+i\pi \Theta(-\rho_{{\rm B}})\right] +\int_{\omega_{\rm th}(P)}^{+\infty}\!\frac{d\omega'}{\pi}\Im \Gamma^{{\rm R}}({\bf P},\omega')\left[\ln{\vert \rho_{{\rm B}}^{\rm II} \vert}-i\pi \Theta(-\rho_{{\rm B}}^{\rm II})\right]\right\}\nonumber\\
&+&\frac{m_{\rm F}}{k}\!\!\!\int_{k_{\mu_{{\rm F}}}-k}^{k_{\mu_{{\rm F}}}+k}\!\frac{d P P}{4\pi^{2}} \left\{w(P)\left[\ln{\vert \rho_{{\rm B}}^{\rm I} \vert}+i\pi \Theta(-\rho_{{\rm B}}^{\rm I})\right]_{\omega'=\omega_{\rm CF}(P)}+\int_{\omega_{\rm th}(P)}^{+\infty}\!\frac{d\omega'}{\pi}\Im \Gamma^{{\rm R}}({\bf P},\omega')\left[\ln{\vert \rho_{{\rm B}}^{\rm I} \vert}+i\pi \Theta(-\rho_{{\rm B}}^{\rm I})\right]\right\}\!\!,
\end{eqnarray}
where $\rho_{{\rm B}}^{\rm II}=(\omega-\omega'+\xi^{\rm F}_{P-k})/(\omega-\omega' +\xi^{\rm F}_{P+k})$.
\end{widetext}

\subsection{Retarded T-matrix}
The analytic expression for the retarded T-matrix  $\Gamma^{{\rm R}}$ at zero-temperature is given by:
\begin{equation}
\Gamma^{\rm R}({\bf P},\omega)=- \left[ \frac{m_{r}}{2\pi a}
+i\frac{m_r}{2\pi} k_\omega-I_{\rm F}(P, \omega)\right]^{-1}\!\!\!\!,
\label{gammazero}
\end{equation}
where $k_\omega \equiv \sqrt{2 m_r \left(\mu_{\rm B}+\mu_{\rm F}+ \omega-\frac{P^2}{2M}\right)}$, while the term $I_{\rm F}$ is given by
\begin{eqnarray}
&&\!\!\!\!I_{\rm F}(P, \omega)=\frac{m_{\rm B} \left(k_{\mu_{\rm F}}^2-k_{P}^2-k^2_\omega\right)}{8 \pi^2 P}
\ln\left[ \frac{(k_{\mu_{\rm F}}+k_{P})^2-k_\omega^2}{(k_{\mu_{\rm F}}-k_{P})^2-k_\omega^2}\right]\nonumber\\
&&\!\!\!\!- \frac{m_r k_\omega}{4\pi^2}\left\{
 \ln\left[\frac{(k_{\mu_{\rm F}}+k_\omega)^2-k_{P}^2}{k_{P}^2-(k_{\mu_{\rm F}}-k_\omega)^2}\right]-i \pi \right\}+\frac{m_r k_{\mu_{\rm F}}}{2\pi^2} \label{If}
\end{eqnarray}
where $k_{\mu_{\rm F}}\equiv\sqrt{2 m_{\rm F}\mu_{\rm F}}$, $k_{P} \equiv \frac{m_{\rm F}}{M} P$, and the principal branches 
of the square-root and logarithm functions are meant in the above equations.

\subsection{Threshold frequencies}
We report here the threshold frequencies delimiting the different continuous bands of the fermionic and bosonic spectral functions. 
For convenience we repeat here also some threshold frequencies that where scattered through the text. For the fermionic spectral function, there is just one
continuous band which is limited from below  by $ \omega^{<}(k)=-\xi^{\rm B}_{k+k_{\rm CF}}$ and from above by
\begin{equation}
\omega^{>}(k)=
\begin{cases}
\frac{k^2}{2(M^*-m_{\rm B})}-\mu^*_{\rm CF}+\mu_{\rm B}& \phantom{aa}  \frac{k}{k_{\rm CF}}  < 1- \frac{m_{\rm B}}{M^*}\\
-\xi^{\rm B}_{k-k_{\rm CF}} &   \phantom{aa}\frac{k}{k_{\rm CF}} \geq 1- \frac{m_{\rm B}}{M^*} \; .
\end{cases}
\end{equation}
For the bosonic spectral function we distinguish in general three bands. 

The uppermost band, which originates from the continuum part of the composite-fermion spectral function $\Im \Gamma^{\rm R}({\bf P},\omega)$, is limited from below by $-\mu_{\rm B}$ for $k \le 2 k_{\mu_{\rm F}}$ or by $\omega_{\rm th}(k-k_{\mu_{\rm F}})$ for $k > 2k_{\mu_{\rm F}}$ (before it merges with the intermediate band). 

For the intermediate band  we distinguish two cases: {\em (a)} If $k_{\mu_{\rm F}} \le k_{\rm CF}$ (a situation which occurs for small density imbalance),  the intermediate band is absent for  $k< k_{\rm CF}-k_{\mu_{\rm F}}$. It is limited from below 
by 
\begin{equation}
\omega_{+}^{<}(k)= 
\begin{cases}
0 &k_{\rm CF} -k_{\mu_{\rm F}}\le k \leq   k_{\mu_{\rm F}}+k_{\rm CF}\\
\omega_{\rm CF}(k-k_{\mu_{\rm F}}) &k_{\mu_{\rm F}}+k_{\rm CF}<k \;.
\end{cases}
\end{equation}
 For the upper limiting frequency  $\omega_{+}^{>}(k)$ we need to distinguish two  further sub-cases.
 If $k_{\rm CF}\frac{m_{\rm F}}{M^*}<k_{\mu_{\rm F}}$ then
\begin{equation}
\omega_{+}^{>}(k)=
\begin{cases}
-\xi^{\rm F}_{k-k_{\rm CF}}  &k_{\rm CF} -k_{\mu_{\rm F}}\le k \leq k_1\\
\frac{k^2/2}{M^*-m_{\rm F}}-\mu^*_{\rm CF}+\mu_{\rm F}&k_1<k \le k_2\\
\omega_{\rm CF}(k+k_{\mu_{\rm F}}) & k_2 < k\;,
\end{cases}
\end{equation}
 where
 \begin{eqnarray}
 k_1&=&k_{\rm CF}(1-m_{\rm F}/M^*)\\
k_2&=&k_{\mu_{\rm F}}(M^*/m_{\rm F}-1) ,
\end{eqnarray}
while for $k_{\rm CF}\frac{m_{\rm F}}{M^*} > k_{\mu_{\rm F}}$ 
\begin{equation}
\omega_{+}^{>}(k)=
\begin{cases}
-\xi^{\rm F}_{k-k_{\rm CF}}  
&k_{\rm CF} -k_{\mu_{\rm F}}\le k \leq k_2\\
\omega_{\rm CF}(k+k_{\mu_{\rm F}}) 
& k_2 < k\;.
\end{cases}
\end{equation}
For both sub-cases, the intermediate band will eventually merge with the uppermost band at sufficiently large $k$.
 {\em (b)} For $k_{\mu_{\rm F}} > k_{\rm CF}$ (i.e., at at large density imbalance), the intermediate band is present for all values of $k$ (before it merges with the uppermost band at large $k$) and is limited from below by
 \begin{equation}
\omega_{+}^{<}(k)= 
\begin{cases}
\omega_{\rm CF}(k-k_{\mu_{\rm F}}) &k\le k_{\mu_{\rm F}}-k_{\rm CF}\\
0 &k_{\mu_{\rm F}}-k_{\rm CF} < k \le   k_{\mu_{\rm F}}+k_{\rm CF}\\
\omega_{\rm CF}(k-k_{\mu_{\rm F}}) &k_{\mu_{\rm F}}+k_{\rm CF}<k \;,
\end{cases}
\end{equation}
 while for the upper frequency we have
\begin{equation}
\omega_{+}^{>}(k)=
\begin{cases}
-\xi^{\rm F}_{k-k_{\rm CF}}  &k \leq k_1\\
\frac{k^2/2}{M^*-m_{\rm F}}-\mu^*_{\rm CF}+\mu_{\rm F}&k_1<k \le k_2\\
\omega_{\rm CF}(k+k_{\mu_{\rm F}}) & k_2 < k\;,
\end{cases}
\end{equation}
before it merges with the uppermost band.

Also for the lowermost band  we need to distinguish the same two cases. 

{\em (a)}  If $k_{\mu_{\rm F}}\ge k_{\rm CF}$, the lowermost band is present only for 
 $k> k_{\mu_{\rm F}}-k_{\rm CF}$ and is limited from below by $\omega_{-}^{<}(k)=-\xi^{\rm F}_{k+k_{\rm CF}}$ and 
 from above by
 \begin{equation}
\omega_{-}^{>}(k)=
\begin{cases}
0&  k_{\mu_{\rm F}}-k_{\rm CF} \le k  \le k_{\mu_{\rm F}}+k_{\rm CF}\\
 -\xi^{\rm F}_{k-k_{\rm CF}}  & k_{\mu_{\rm F}}+k_{\rm CF}< k\,.
\end{cases}
\end{equation}

{\em (b)} If  $k_{\mu_{\rm F}}< k_{\rm CF}$, the lowermost band is present for all values of $k$ and is again limited from below by $\omega_{-}^{<}(k)=-\xi^{\rm F}_{k+k_{\rm CF}}$.
For the limiting frequency from above, we need to distinguish the same two sub-cases considered above for the intermediate band.
For $k_{\rm CF}\frac{m_{\rm F}}{M^*} < k_{\mu_{\rm F}}$ 
\begin{equation}
\omega_{-}^{>}(k)=
\begin{cases}
{\rm min}\!\left[\frac{k^2/2}{M^*-m_{\rm F}}-\mu^*_{\rm CF}+\mu_{\rm F},0\right]  & k \le k_{\rm CF}-k_{\mu_{\rm F}}\\
0  & \!\!\!\!\!\!\!\!\!\!\!\!\!\!\!\!\!\!\!\!\!\!\!\!\!\!\!\!\!\!\!\!\!\!k_{\rm CF}-k_{\mu_{\rm F}}< k \le k_{\mu_{\rm F}}+k_{\rm CF}\\
-\xi^{\rm F}_{k-k_{\rm CF}} &k \geq k_{\mu_{\rm F}}+k_{\rm CF}
\end{cases}
\end{equation}
 while for $k_{\rm CF}\frac{m_{\rm F}}{M^*} > k_{\mu_{\rm F}}$
 \begin{equation}
\omega_{-}^{>}(k)=
\begin{cases}
\frac{k^2/2}{M^*-m_{\rm F}}-\mu^*_{\rm CF}+\mu_{\rm F} \phantom{1234567891234567}k \le k_1&\\
-\xi^{\rm F}_{k-k_{\rm CF}} \phantom{123456789123456}k_1< k \le k_{\rm CF}-k_{\mu_{\rm F}}&\\
0  \phantom{aaaaaaaaaaaaa!}k_{\rm CF}-k_{\mu_{\rm F}}< k \le k_{\mu_{\rm F}}+k_{\rm CF}&\\
-\xi^{\rm F}_{k-k_{\rm CF}}  \phantom{aaaaaaaaaaaaaaaaaa!} k_{\mu_{\rm F}}+k_{\rm CF} < k .&
\end{cases}
\end{equation}

\end{document}